\documentclass{llncs} %
\usepackage{filecontents,catchfile}
\usepackage{amsmath,amssymb,amsfonts}
\usepackage{multirow}
\usepackage{multicol}
\usepackage{graphicx}
\usepackage{lscape} 
\usepackage[table]{xcolor} 
\usepackage{array}

\usepackage{float}
\usepackage{wrapfig}
\usepackage[draft]{todonotes}   % notes shown
\usepackage[labelfont=bf]{caption}
\usepackage{rotating}
\usepackage{caption}
\usepackage{mathtools} % Mathe

\usepackage{booktabs}
\usepackage{rotating}

\usepackage[export]{adjustbox}

\usepackage{color}

\definecolor{azure}{rgb}{0.0, 0.5, 1.0}

\usepackage[draft]{todonotes}   % notes shown
 
\usepackage{soul}

\definecolor{darkgreen}{rgb}{0.53, 0.66, 0.42}
\definecolor{grey}{rgb}{0.57, 0.64, 0.69}
\usepackage[colorlinks=true,linkcolor=blue,citecolor=blue]{hyperref} 

\usepackage{float}
\usepackage{graphicx}
\usepackage{wrapfig}
\usepackage{placeins}
\usepackage[labelfont=bf]{caption}
\DeclareMathAlphabet\mathbfcal{OMS}{cmsy}{b}{n}
\usepackage{color,soul}
\usepackage{amsmath} % Mathe
\usepackage{mathtools} % Mathe
\usepackage{amsfonts} % Mathesymbole
\usepackage{amssymb}

\usepackage{array}
\usepackage{booktabs}
\usepackage{rotating}
\usepackage{multirow}
\usepackage{multicol}
\usepackage{lscape}
\usepackage{subfig}

\usepackage[export]{adjustbox}

\usepackage{algorithm,algorithmicx,algpseudocode}

\usepackage[labelfont=bf]{caption}
\usepackage[colorlinks=true,linkcolor=blue,citecolor=blue]{hyperref}

\usepackage{color}

\usepackage{xargs}                      % Use more than one optional parameter in a new commands

\usepackage{color}

\definecolor{darkgreen}{rgb}{0.53, 0.66, 0.42}
\usepackage{tikz,xcolor,hyperref}

\definecolor{lime}{HTML}{A6CE39}
\DeclareRobustCommand{\orcidicon}{
	\begin{tikzpicture}
	\draw[lime, fill=lime] (0,0) 
	circle [radius=0.16] 
	node[white] {{\fontfamily{qag}\selectfont \tiny ID}};
	\draw[white, fill=white] (-0.0625,0.095) 
	circle [radius=0.007];
	\end{tikzpicture}
	\hspace{-2mm}
}

\foreach \x in {A, ..., Z}{\expandafter\xdef\csname orcid\x\endcsname{\noexpand\href{https://orcid.org/\csname orcidauthor\x\endcsname}
			{\noexpand\orcidicon}}
}
			
\begin{document}

\title{A Few-shot Learning Graph Multi-Trajectory Evolution Network for Forecasting Multimodal Baby Connectivity Development from a Baseline Timepoint}

\titlerunning{Short Title}  % abbreviated title (for running head)

\author{Alaa Bessadok\orcidB{}\index{Bessadok, Alaa}\inst{1,2,3} \and Ahmed Nebli\orcidC{}\index{Nebli, Ahmed}\inst{1,4} \and Mohamed Ali Mahjoub\index{Mahjoub, Mohammed Ali}\inst{2,3}  \and Gang Li\orcidF{} \index{Li, Gang}\inst{5}  \and Weili Lin\orcidD{}\index{Lin, Weili}\inst{5} \and  Dinggang Shen\orcidE{}\index{Shen, Dinggang}\inst{6,7} \and Islem Rekik\orcidA{}\index{Rekik, Islem}\inst{1}\thanks{ {corresponding author: irekik@itu.edu.tr, \url{http://basira-lab.com}. }}}

\institute{$^{1}$ BASIRA Lab, Faculty of Computer and Informatics, Istanbul Technical University, Istanbul, Turkey \\ $^{2}$ Higher Institute of Informatics and Communication Technologies, University of Sousse, Tunisia, 4011 \\ $^{3}$ National Engineering School of Sousse, University of Sousse, LATIS- Laboratory of Advanced Technology and Intelligent Systems, Sousse, Tunisia, 4023 \\  $^{4}$ National School for Computer Science (ENSI), Mannouba, Tunisia \\ $^{5}$ Department of Radiology and BRIC, University of North Carolina at Chapel Hill, NC, USA \\ $^{5}$ School of Biomedical Engineering, ShanghaiTech University, Shanghai, China \\ $^{7}$ Department of Research and Development, United Imaging Intelligence Co., Ltd., Shanghai, China}

\authorrunning{A. Bessadok et al.}

\maketitle              % typeset the title of the contribution

\begin{abstract}

Charting the baby connectome evolution trajectory during the first year after birth plays a vital role in understanding dynamic connectivity development of baby brains. Such analysis requires acquisition of longitudinal connectomic datasets. However, both neonatal and postnatal scans are rarely acquired due to various difficulties. A small body of works has focused on predicting baby brain evolution trajectory from a neonatal brain connectome derived from a single modality. Although promising, large training datasets are essential to boost model learning and to generalize to a multi-trajectory prediction from different modalities (i.e.,  functional and morphological connectomes). Here, we unprecedentedly explore the question: \emph{``Can we design a few-shot learning-based framework for predicting brain graph trajectories across different modalities?"} To this aim,  we propose a Graph Multi-Trajectory Evolution Network (GmTE-Net), which adopts a teacher-student paradigm where the teacher network learns on pure neonatal brain graphs and the student network learns on simulated brain graphs given a set of different timepoints. To the best of our knowledge, this is the first teacher-student architecture tailored for brain graph multi-trajectory growth prediction that is based on few-shot learning and generalized to graph neural networks (GNNs). To boost the performance of the student network, we introduce a local topology-aware distillation loss that forces the predicted graph topology of the student network to be consistent with the teacher network. Experimental results demonstrate substantial performance gains over benchmark methods. Hence, our GmTE-Net can be leveraged to predict atypical brain connectivity trajectory evolution across various modalities. Our code is available at \url{https://github.com/basiralab/GmTE-Net}.

\end{abstract}

\keywords{Few-shot learning using graph neural network $\cdot$ Multimodal brain graph evolution prediction $\cdot$ Baby connectome development $\cdot$ Knowledge distillation network}

%% ***************************************************************************** %%
\section{Introduction}
%% ***************************************************************************** %% 

\begin{figure}[ht!]
\centerline{\includegraphics[width=13cm]{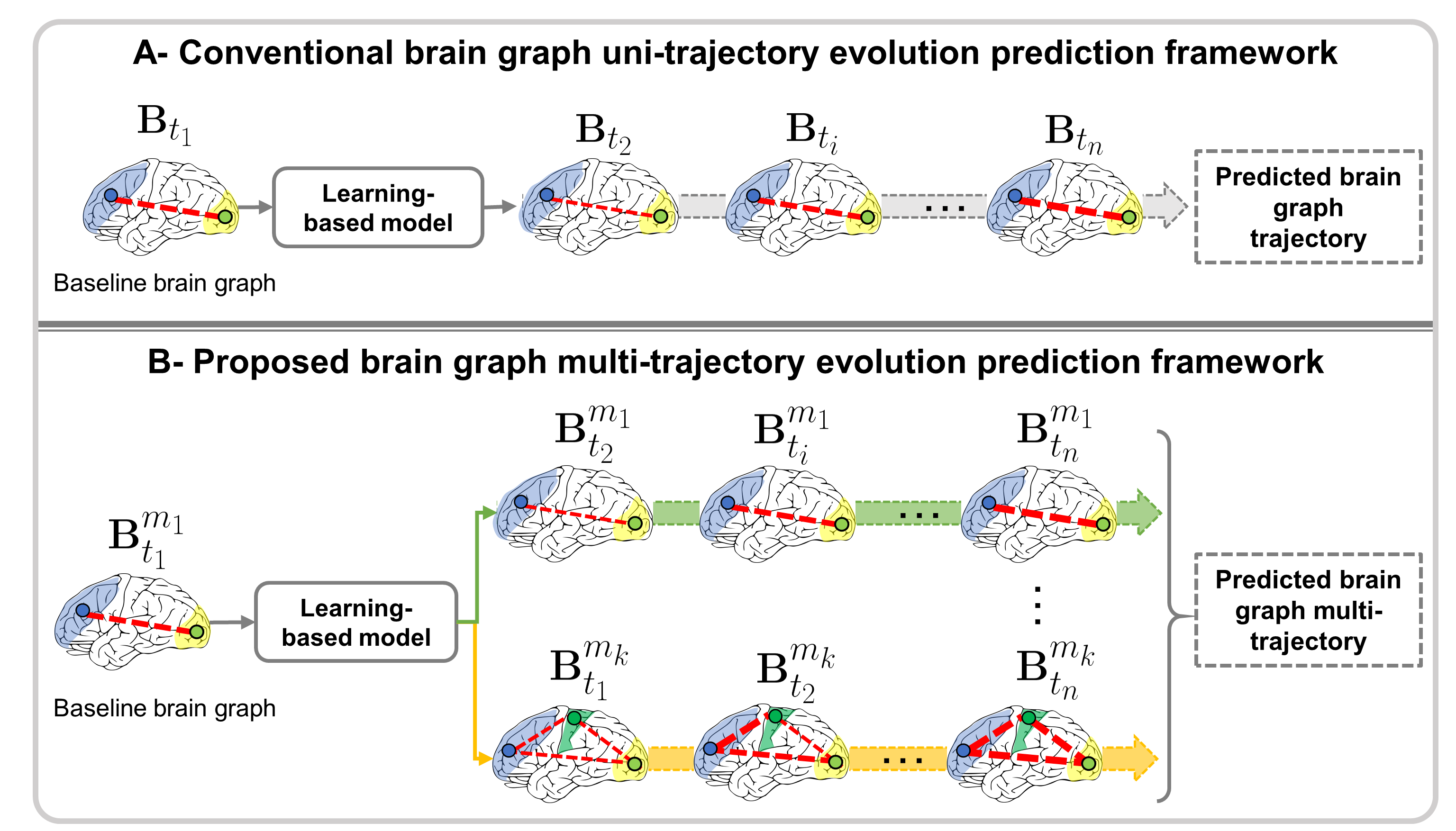}}
\caption{\emph{Conventional brain graph uni-trajectory evolution prediction methods and the proposed brain graph multi-trajectory evolution prediction architecture.}}% 
\label{1}
\end{figure}

During the first year after birth, the baby brain undergoes the most critical and dynamic postnatal development in structure, function, and morphology \cite{zhang2019resting}. Such dramatic changes are highly informative for future brain diseases that can be easily preventable if \emph{predicted} at an early stage \cite{rekik2016predicting,rekik2017joint}.  However, this problem is challenging due to the scarcity of longitudinal datasets. Thus, there is an increasing need to foresee longitudinal missing baby connectomic data given what already exists \cite{GNNreviewAlaa}. Network literature \cite{fornito2015connectomics} shows that a typical brain connectome is defined as a graph of brain connectivities where nodes denote regions of interest (ROIs) and edges encode the pairwise connectivity strengths between ROI pairs. As such, \cite{ghribi2021multi} attempted to predict baby brain evolution trajectory from a single neonatal brain graph. The authors designed an ensemble of regressors for a sample selection task, thereby progressively predicting a follow-up baby brain connectome from a single timepoint. While effective, this method is limited in the following ways: (i) the learning model lacks preservation of topological properties of the original brain graphs, and (ii) the framework is not designed in an end-to-end learning manner which may lead to accumulated errors across the learning steps. To overcome these limitations, \cite{goktas2020residual} proposed a model which learns an adversarial brain graph embedding using graph convolution networks (GCN) \cite{Kipf:2016} and introduced a sample selection strategy guided by a population brain \emph{template}. Another work also proposed a infinite sample selection strategy for brain network evolution prediction from baseline \cite{ezzine2019learning}. However, both models were dichotomized into subparts during the learning step. In another study, \cite{nebli2020deep} proposed an end-to-end learning framework based on Generative Adversarial Network (GAN) that successfully predicts the follow-up brain graphs in a cascaded manner. However, such design choice exclusively predicts \emph{uni-trajectory} evolution where brain graphs are derived from a single modality, limiting its generelizability to \emph{multi-trajectory} graph evolution prediction where each trajectory represents a different brain modality (i.e., structural and functional connectomes) (\textbf{Fig.}~\ref{1}).

A second major limitation is requiring many training brain graphs to boost model learning, reducing its learning capacity in a frugal setting where only \emph{a few} training samples exist. Specifically, due to the high medical image acquisition costs, it is often challenging to acquire all modalities for each subject. Hence, it is mandatory to learn how to predict brain graphs from a small number of graphs effectively. This has initiated a new line of research, namely few-shot learning, which aims to bridge the gap between the breadth of deep learning algorithms and the necessity to use small data sets \cite{tian2020few,li2019large}. In other words, few-shot learning aims to extract the most relative information from each training data point without the need for acquiring a large number of instance.

In this context,  several works have demonstrated the feasibility of learning given few samples \cite{li2020difficulty,yuan2020few}. However, none of these addressed the problem of multi-trajectory evolution prediction. To tackle these limitations, \textbf{we propose Graph Multi-Trajectory Evolution Network (GmTE-Net), the first few-shot learning framework designed for baby brain graph multi-trajectory evolution prediction using a teacher-student (TS) paradigm \cite{hinton2015distilling}.} Namely, TS is a set of two networks where an extensive network called teacher aims to transfer its knowledge to a smaller network called student. Intuitively, if the teacher network can efficiently transfer its knowledge to the student network, then we can affirm the high learning quality of the teacher, and we can condition the learning curve of the student. Such architecture is highly suitable for few-shot learning \cite{rajasegaran2020self} thanks to its high ability to generalize to different data distributions and enforce robustness against perturbations.

Also, unlike \cite{nebli2020deep}, this architecture breaks free from the cascaded scheme of stacked networks enabling prediction to go ``multi-trajectory". Our model aims to ensure the biological soundness of the predicted brain graphs by defining a \emph{global topology loss} to train the teacher network. For the student network, we propose a novel \emph{local topology-aware distillation loss} which enforces the student model to capture the potentially complex dynamics of the local graph structure generated by the teacher over time. This enables topology-aware knowledge transfer from the teacher, yielding a compact yet high-performance student model. Ultimately, we present the multi-level contributions of our work:

\begin{enumerate}

\item \emph{On a conceptual level.} Our proposed GmTE-Net is the first teacher-student framework tailored for jointly predicting multiple trajectories of infant brain graphs from a single neonatal observation.

\item \emph{On a methodological level.} GmTE-Net is a novel knowledge distillation approach where the student model benefits from the few-shot learning of the teacher model. We also propose a local topology-aware distillation loss to transfer the local graph knowledge from the teacher to the student. We further propose a global topology loss to optimize the learning of the teacher network.

\item \emph{On clinical level.} GmTE-Net provides a great opportunity for charting the rapid and dynamic brain developmental trajectories of baby connectomes which help to spot early neurodevelopmental disorders e.g., autism.

\end{enumerate}

\begin{figure}[htp!]
\hspace{-40pt}
\includegraphics[width=15cm]{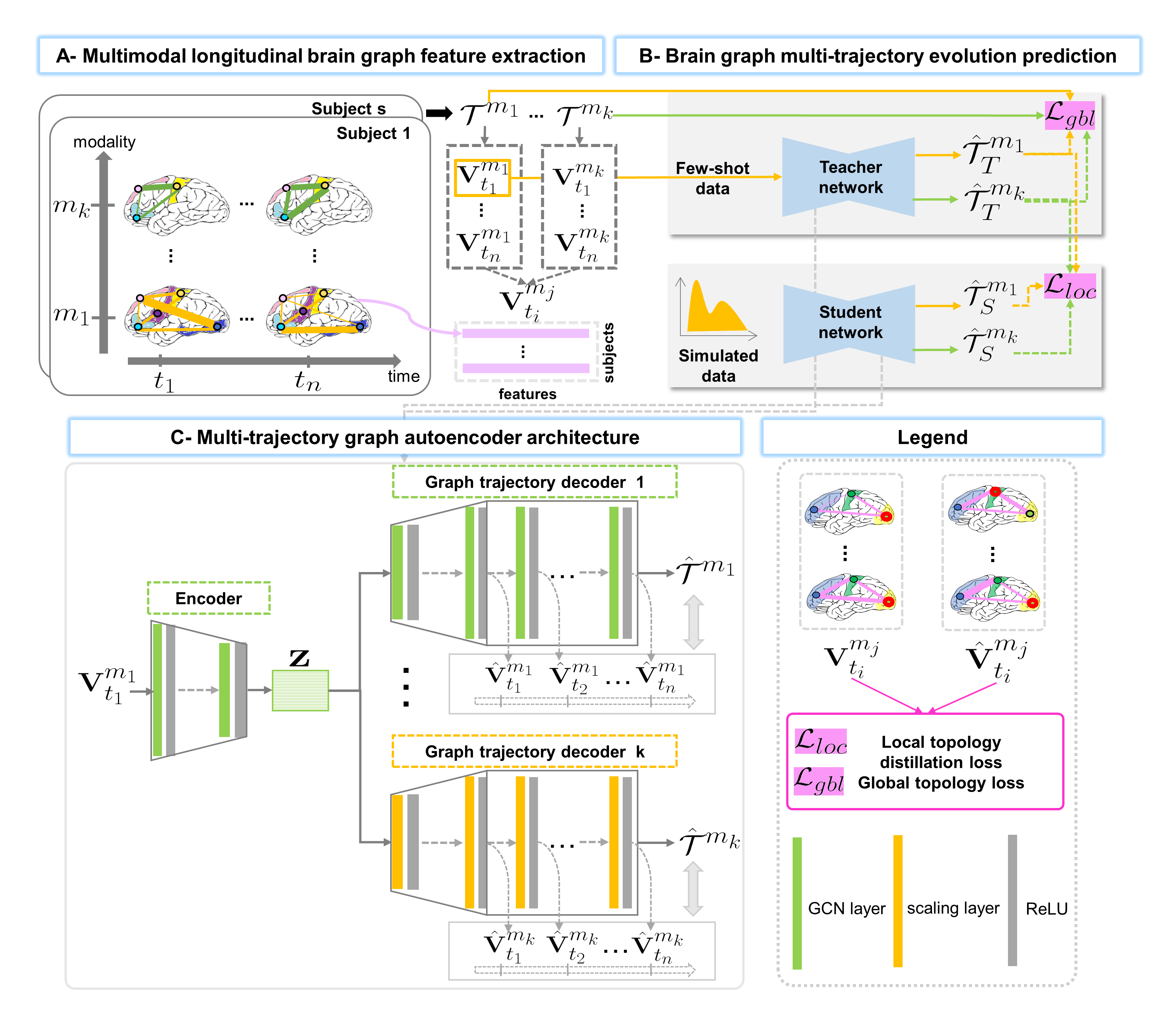}
\caption{\emph{Pipeline of the proposed GmTE-Net framework for predicting multi-trajectory evolution of baby brain graphs from a single neonatal brain graph.} \textbf{(A) Multimodal longitudinal brain graph feature extraction.} Each subject is represented by $n_m$ trajectories (i.e., $\mathcal{T}^{m_j}$), each representing a set of brain graphs derived from multiple modalities at a specific timepoint. Since each brain graph is encoded in a symmetric matrix, we vectorize the off-diagonal upper-triangular part and stack the resulted vectors in a matrix $\mathbf{V}^{m_j}_{t_i}$ for each brain modality. \textbf{(B) Brain multi-trajectory evolution prediction.} We first train a teacher network with a few-shot learning strategy. Given a feature matrix $\mathbf{V}^{m_1}_{t_1}$ representing a baseline timepoint, we aim to predict multi-trajectory evolution where each trajectory is a set of follow-up brain graphs of a specific modality (i.e., $\hat{\mathcal{T}}_{T}^{m_j}$). Second, we propose to train a student network on simulated data to boost the generalization capability in learning from any data distribution. We regularize the teacher network with a \emph{global topology loss} and the student network with a \emph{local topology-aware distillation loss} so that they capture both global and local node properties. \textbf{(C) Multi-trajectory graph autoencoder architecture.} Both teacher and student are GCN-based graph autoencoders encapsulating an encoder and a set of decoders each aiming to predict a graph trajectory for a specific modality. }
\label{2}
\end{figure}

% %% ***************************************************************************** %%
\section{GmTE-Net for Brain Graph Multi-Trajectory Evolution Prediction}
% %% ***************************************************************************** %%

\textbf{Problem Definition.} A brain can be represented as a graph \( \mathcal{B}=\{ \mathbf{R}, \mathbf{E}, \mathbf{v} \}\) where \(\mathbf{R}\) is a set of nodes (i.e., brain ROIs) and \(\mathbf{E}\) is a set of weighted edges. $\mathbf{v}$ is a feature vector denoting a compact and reduced representation of the brain connectivity matrix measuring the pairwise edge weight between nodes. Each baby sample $s$ is represented by a set of brain graph \emph{multi-trajectories}  $\mathcal{T}^s=\{\mathcal{T}^{m_j}\}_{j=1}^{n_m}$ where $s \in \{1, \dots, n_s \}$ and $n_m$ is the number of modalities (e.g., functional, morphological). A single trajectory derived from the $m_j$-th modality can be written as $\mathcal{T}^{m_j}=\{\mathbf{v}^{m_j}_{t_i}\}_{i=1}^{n_t} $ where $\mathbf{v}^{m_j}_{t_i}$ denotes brain feature vector at the $i$-th timepoint $t_i$, where $i \in \{1, \dots, n_t \}$. Given a baseline testing brain feature vector $\mathbf{v}^{m_1}_{t_1}$ derived from a single modality $m_1$, we aim to predict its evolution across different $n_m$ trajectories: one \emph{base} trajectory derived from $m_1$ and other $\{n_m - 1\}$ trajectories derived from different modalities. Note that for the base trajectory we only need to predict the evolving graph for $t \in \{2, \dots, n_t \}$ unlike the other trajectories $\{\{\hat{\mathbf{v}}^{m_j}_{t_i}\}_{i=1}^{n_t}\}_{j=2}^{n_m}$ where the neonatal observation is also missing. We note that for simplicity, $k=n_m$ and $n=n_t$ in \textbf{Fig.}~\ref{1} and \textbf{Fig.}~\ref{2}.

\textbf{A- Multimodal longitudinal brain graph feature extraction}. The ultimate goal of this work is to predict a set of evolution trajectories given a single brain graph, each mapping a particular brain modality. Since a brain graph is encoded in a symmetric matrix composed of weighted edge values, we propose to vectorize its off-diagonal upper-triangular part. We stack the resulting feature vectors of all subjects into a matrix $\mathbf{V}^{m_j}_{t_i}$ representing the brain features derived from the modality $m_j$ at timepoint $t_i$.We repeat the same step for each timepoint to obtain a tensor $\mathcal{T}^{m_j}=\{\mathbf{V}^{m_j}_{t_i}\}_{i=1}^{n_t}$. For our training set, we obtain $n_m$ \emph{multi-trajectory} evolution $\mathcal{T}=\{\mathcal{T}^{m_j}\}_{j=1}^{n_m}$ representing all training subjects (\textbf{Fig.}~\ref{2}--A).

\textbf{B- Brain multi-trajectory evolution prediction.} Owing to the existing few-shot learning works \cite{yuan2020few,li2020difficulty}, it has been demonstrated that training a deep learning framework using limited data is feasible, but it never outperforms same models with large data. Under this assumption,  training a single network with a few brain graphs cannot be efficient for predicting brain \emph{multi-trajectory} evolution.  Thus, we propose a TS scheme where we first train a teacher network using a limited number of brain graphs so that it learns the \emph{pure} infant connectivity patterns. Second, we freeze the teacher model and train a student network on \emph{simulated} data such that the network can generalize to different types of brain connectomes. To simulate the longitudinal connectomic data, we generate a random dataset having the same distribution properties (i.e., mean and standard deviation) of the pure neonatal brain graph data. We design both teacher and student models as two graph autoencoders; each composed of an encoder $E$ aiming to learn the latent representation from the baseline brain graph and a set of $\{D^{m_j}\}_{j=1}^{n_m}$ \emph{graph trajectory} decoders. While the vanilla TS framework \cite{hinton2015distilling} consists in a larger teacher network and a shallow student network, we choose to design both networks with identical encoder and decoder architectures which has been proved to be effective in many tasks \cite{hu2020knowledge,zhou2020deep}. We define the encoder and decoder networks as follows:

\begin{enumerate}

    \item \textbf{\emph{The encoder.}} $E(\mathbf{V}^{m_j}_{t_i}, \mathbf{A})$ aims to map a set of input brain graphs into a low-dimensional space. To do so, our encoder takes as an input $\mathbf{V}^{m_j}_{t_i}$, the feature matrix stacking brain vectors of samples at the initial timepoint ${t_i}$ for ${m_j}$ and an adjacency matrix $\mathbf{A}$. We initialize the adjacency matrix as an identity one to (i) eliminate redundancy of edge weights and (ii) compensate for the vectorized edge features. The proposed encoder architecture is composed of two layers, each containing a GCN layer, followed by a Rectified Linear Unit (ReLU) and a dropout function.

    \item \textbf{\emph{The decoder.}} $D(\mathbf{Z}^{l}, \mathbf{A})$ aims to predict a modality-specific trajectory $m_j$ for the follow-up brain graphs. To do so, our decoder takes as input the latent space $\mathbf{Z}^{l}$, and an identity matrix $\mathbf{I}$. Similarly to the encoder,  we define the decoder as a two-layer GCN, each followed by a ReLU and a dropout functions. This elementary GCN-based architecture is cascaded $n_t$ times where each predicts a brain feature matrix at a different timepoint. Specifically, the first GCN layer might play a scaling role in case of graph resolution (i.e., node size) difference between initial and target domains. In such scenario, it maps the base brain graph to the follow-up trajectory-specific resolutions. As so, our decoder is able to predict brain graphs even if there are resolution shifts across timepoints.

\end{enumerate}

Our proposed autoencoder is capable of foreseeing multiple trajectories given a single brain graph modality acquired at the initial timepoint. \emph{In essence, from a single shared low-dimensional space of baseline brain graphs it charts out the follow-up brain graphs for different trajectories and with potential resolution shifts across trajectories.} We define the propagation rule of our GCN-based autoencoder as follows:

%\centerline{$
\begin{equation*}
    \mathbf{Z}^{(l)} = f_{ReLU}(\mathbf{F}^{(l)}, \mathbf{A} \vert \mathbf{W}^{(l)}) = {ReLU}(\mathbf{\widetilde{D}}^{-\frac{1}{2}}\mathbf{\widetilde{\mathbf{A}}}\mathbf{\widetilde{D}}^{-\frac{1}{2}}\mathbf{F}^{(l)}\mathbf{W}^{(l)})
\end{equation*}

$\mathbf{Z}^{(l)}$ is the learned graph representation resulting from the layer $l$.  For the encoder, $\mathbf{F}^{(l)}$ is defined as $\mathbf{V}^{m_j}_{t_i}$ in the first layer and as the learned embeddings $\mathbf{Z}^{l}$ in the second layer. We define the output of the decoder for $n_t$ layers as the trajectory $\mathcal{T}^{m_j}=\{\mathbf{V}^{m_j}_{t_i}\}_{i=1}^{n_t}$.  $\mathbf{W}^{(l)}$ is a filter denoting the graph convolutional weights in layer $l$. We define the graph convolution function as $f_{(.)}$ where $\mathbf{\widetilde{\mathbf{A}}} = \mathbf{\mathbf{A}} + \mathbf{I}$ with $\mathbf{I}$ being an identity matrix, and $\mathbf{\widetilde{D}}_{ii} = \sum_{j}\mathbf{\widetilde{\mathbf{A}}_{ij}}$ is a diagonal matrix (\textbf{Fig.}~\ref{2}--C). 

Despite the effective graph representation learning of GCN, it might fail in preserving both global and local graph topological structures. To enforce the teacher model to capture the high-order representation of input brain graphs, we introduce a \emph{global topology loss} $ \mathcal{L}_{glob}^{m_j}$ defined as the average mean absolute error (MAE) between the ground truth brain features $\mathbf{V}^{m_j}_{t_i}$ and the predicted ones $\hat{\mathbf{V}}^{m_j}_{t_i}$ over timepoints. This loss aims to learn the global graph structure given few examples per modality, thereby synthesizing biologically sound brain graph \emph{multi-trajectory} predictions. We define the teacher loss function and our \emph{global topology loss} as follows:
\begin{equation*}
 	\mathcal{L}_{Teacher} = mean(\sum_{j=1}^{n_m} \mathcal{L}_{glob}^{m_j});
    \quad\text{ }
    \mathcal{L}_{glob}^{m_j} = mean( \sum_{i=1}^{n_t} \ell_{MAE}(\mathbf{V}^{m_j}_{t_i}, \hat{\mathbf{V}}^{m_j}_{t_i}))
\label{eq:10}
\end{equation*}
Following the teacher training, we freeze it in order to train the student model. To generalize the student network to be distribution agnostic, we train it on a large set of simulated data. To this end, it is difficult to assume node property preservation for the predicted brain graphs with respect to the ground-truth training. Interestingly, brain graphs have unique topological properties for functional, structural and morphological connectivities \cite{Bassett:2017} that should be preserved when synthesizing the brain graphs of a specific trajectory \cite{Liu:2017,Joyce:2010}. Thus, we introduce a \emph{local topology-aware distillation loss} that aims to force the student network to mimic the local topology of the predicted graphs from the teacher model. As so, we are training the student using the local topological knowledge distilled from the teacher network. We denote our loss by $ \mathcal{L}_{loc}^{m_j}$ defined as the MAE between the centrality scores of the ground truth and the predicted graphs. We choose the closeness centrality measure since it is mostly used in graph theory \cite{fornito2016fundamentals}. It quantifies the closeness of a node to all other nodes \cite{Freeman:1977} which is defined by the following formula: $C(r^{a}) = \frac{r-1}{\sum_{r^{a}\neq r^{b}} p_{r^{a}r^{b}}}$, where $r$ denotes the number of nodes and $p_{r^{a}r^{b}}$ the length of the shortest path between nodes $r^{a}$ and $r^{b}$. So, we first create a centrality vector for each brain graph where its elements represent the computed centrality scores for the graph nodes. Then, we vertically stack the resulting vectors and create a centrality matrices $\mathbf{C}$ and $\hat{\mathbf{C}}$ for ground-truth and predicted brain graphs, respectively. Therefore, we define the student loss function and the \emph{local topology-aware distillation loss} as follows:
\begin{equation*}
    \mathcal{L}_{Student} = mean(\sum_{j=1}^{n_m} \mathcal{L}_{loc}^{m_j});
    \quad\text{ }
    \mathcal{L}_{loc}^{m_j} = mean( \sum_{i=1}^{n_t} \ell_{MAE}(\mathbf{C}^{m_j}_{t_i}, \hat{\mathbf{C}}^{m_j}_{t_i}))
\label{eq:11}
\end{equation*}
In that way, the student network is not only preserving the local topology of the brain graphs but also their global topology yielding an accurate student model. 

% \usepackage{multirow}
\begin{table}[b!]
\centering
\scalebox{0.84}{
\begin{tabular}{
>{\columncolor[HTML]{FFFFFF}}c |
>{\columncolor[HTML]{FFFFFF}}c |
>{\columncolor[HTML]{FFFFFF}}c 
>{\columncolor[HTML]{FFFFFF}}c 
>{\columncolor[HTML]{FFFFFF}}c |
>{\columncolor[HTML]{FFFFFF}}c 
>{\columncolor[HTML]{FFFFFF}}c 
>{\columncolor[HTML]{FFFFFF}}c }
\hline
\multicolumn{2}{c}{\cellcolor[HTML]{FFFFFF}{\color[HTML]{000000} }}                                                          & \multicolumn{3}{|c}{\cellcolor[HTML]{FFFFFF}{\color[HTML]{000000} Student evaluation}}                                                                                   & \multicolumn{3}{|c}{\cellcolor[HTML]{FFFFFF}{\color[HTML]{000000} Teacher evaluation}}                                                                                 \\ 
\multicolumn{2}{c|}{\multirow{-2}{*}{\cellcolor[HTML]{FFFFFF}{\color[HTML]{000000} Models}}}                                  & \cellcolor[HTML]{FFCCC9}MAE(graph)      & \cellcolor[HTML]{CBCEFB}MAE(EC)         & \cellcolor[HTML]{FFFFC7}MAE(PC)         & \cellcolor[HTML]{FFCCC9}{\color[HTML]{000000} MAE(graph)} & \cellcolor[HTML]{CBCEFB}MAE(EC)         & \cellcolor[HTML]{FFFFC7}MAE(PC)         \\
\hline\hline
\cellcolor[HTML]{FFFFFF}                                                  & {\color[HTML]{14A4C8} $\mathcal{L}_{a}$} & 0.40821                                 & 0.02948                                 & 0.00393                                 & 0.34529                                                   & 0.02818                                 & 0.00371                                 \\
\cellcolor[HTML]{FFFFFF}                                                  & {\color[HTML]{009901} $\mathcal{L}_{b}$} & 0.48441                                 & 0.0407                                  & 0.00569                                 & 0.48505                                                   & 0.04715                                 & 0.00641                                 \\
\multirow{-3}{*}{\cellcolor[HTML]{FFFFFF}Ablated (augmented + real data)} & {\color[HTML]{A411AA} $\mathcal{L}^{\star}$} & 0.48436                                 & 0.04074                                 & 0.00546                                 & 0.35087                                                   & 0.0284                                  & 0.00377                                 \\\hline
\cellcolor[HTML]{FFFFFF}                                                  & {\color[HTML]{14A4C8} $\mathcal{L}_{a}$} & 0.49334                                 & 0.08182                                 & 0.01016                                 & 0.3481                                                    & {\color[HTML]{14A4C8} \textbf{0.0123}}  & 0.00177                                 \\
\cellcolor[HTML]{FFFFFF}                                                  & {\color[HTML]{009901} $\mathcal{L}_{b}$} & 0.48429                                 & 0.03993                                 & 0.00572                                 & 0.48358                                                   & 0.04024                                 & 0.00566                                 \\
\multirow{-3}{*}{\cellcolor[HTML]{FFFFFF}Ablated (few-shot learning)}     & {\color[HTML]{A411AA} $\mathcal{L}^{\star}$} & 0.48429                                 & 0.03993                                 & 0.00573                                 & 0.34932                                                   & {\color[HTML]{A411AA} \textbf{0.01225}} & {\color[HTML]{A411AA} \textbf{0.00176}} \\\hline\hline
\cellcolor[HTML]{FFFFFF}                                                  & {\color[HTML]{14A4C8} $\mathcal{L}_{a}$} & 0.40856                                 & 0.03032                                 & 0.00404                                 & {\color[HTML]{14A4C8} \textbf{0.31938}}                   & 0.02851                                 & 0.00376                                 \\

\cellcolor[HTML]{FFFFFF}                                                  & {\color[HTML]{009901} $\mathcal{L}_{b}$} & 0.48444                                 & 0.03948                                 & {\color[HTML]{009901} \textbf{0.00555}} & 0.48316                                                   & 0.04779                                 & 0.00666                                 \\
\multirow{-3}{*}{\cellcolor[HTML]{FFFFFF}GmTE-Net}                        & {\color[HTML]{A411AA} $\mathcal{L}^{\star}$} & 0.48413                                 & 0.04666                                 & 0.00625                                 & 0.33871                                                   & 0.02845                                 & 0.00376                                 \\\hline
\cellcolor[HTML]{FFFFFF}                                                  & {\color[HTML]{14A4C8} $\mathcal{L}_{a}$} & {\color[HTML]{14A4C8} \textbf{0.40391}} & {\color[HTML]{14A4C8} \textbf{0.01589}} & {\color[HTML]{14A4C8} \textbf{0.00217}} & 0.32215                                                   & 0.01242                                 & {\color[HTML]{14A4C8} \textbf{0.00174}} \\
\cellcolor[HTML]{FFFFFF}                                                  & {\color[HTML]{009901} $\mathcal{L}_{b}$} & {\color[HTML]{009901} \textbf{0.48369}} & {\color[HTML]{009901} \textbf{0.03875}} & 0.00565                                 & 0.48398                                                   & 0.03997                                 & 0.00522                                 \\
\multirow{-3}{*}{\cellcolor[HTML]{FFFFFF}GmTE-Net$^{\star}$}                       & {\color[HTML]{A411AA} $\mathcal{L}^{\star}$} & {\color[HTML]{A411AA} \textbf{0.48374}} & {\color[HTML]{A411AA} \textbf{0.03846}} & {\color[HTML]{A411AA} \textbf{0.00545}} & {\color[HTML]{A411AA} \textbf{0.3145}}                    & 0.01314                                 & 0.00189      \\\hline\hline                          
\end{tabular}}
\caption{\emph{Morphological connectomic data-based comparison of our GmTE-Net with its two ablated versions in terms of architecture (first two rows) and its two variants (last two rows). We highlight in bold, blue, green and purple the best performances got for each method when using $\mathcal{L}_{a}$, $\mathcal{L}_{b}$ and $\mathcal{L}^{\star}$ and loss functions, respectively.} }
\label{tab:1}
\end{table}

% %% ***************************************************************************** %%
\section{Results and Discussion}
% %% ***************************************************************************** %%

\textbf{Dataset.} We evaluated our framework on 11 baby brain graphs of developing infant subjects each has 4 serial $t_2$-w MRI and resting-state fMRI (rsfMRI) scans acquired at 1, 3, 6, and 9 months of age, respectively. We create for each subject two brain graphs derived from two different brain parcellations: a functional brain graph of size 116$\times$116 using AAL template \cite{tzourio2002automated} and a morphological brain graph of size 35$\times$35 using Desikan–Killiany Atlas \cite{fischl2004sequence}. Out of these 11 subjects, we have four complete (i.e., have all observations and modalities across all timepoints). Using a single Tesla V100 GPU (NVIDIA GeForce GTX TITAN with 32GB memory), we train our teacher network using these four complete subjects and test both teacher and student networks using the whole dataset.

\textbf{Comparison methods and evaluation.} Due to the lack of existing frameworks that aim to predict the multi-trajectory evolution of brain graphs, we benchmark our GmTE-Net against its two ablated versions and two variants, each trained with three different loss functions computed between the predicted and ground truth graphs. Among the three loss functions we have, $\mathcal{L}_{a}$, $\mathcal{L}_{b}$, and  $\mathcal{L}^{\star}$, denoting the MAE loss, the Wasserstein distance, and our proposed loss optimizing the training of both teacher and student networks. We detail our benchmarking methods as follows:

\begin{table}[t!]
\centering
\scalebox{0.84}{
\begin{tabular}{
>{\columncolor[HTML]{FFFFFF}}c |
>{\columncolor[HTML]{FFFFFF}}c |
>{\columncolor[HTML]{FFFFFF}}c 
>{\columncolor[HTML]{FFFFFF}}c 
>{\columncolor[HTML]{FFFFFF}}c |
>{\columncolor[HTML]{FFFFFF}}c 
>{\columncolor[HTML]{FFFFFF}}c 
>{\columncolor[HTML]{FFFFFF}}c }
\hline
\multicolumn{2}{c}{\cellcolor[HTML]{FFFFFF}{\color[HTML]{000000} }}                                                          & \multicolumn{3}{|c}{\cellcolor[HTML]{FFFFFF}{\color[HTML]{000000} Student evaluation}}                                                                                   & \multicolumn{3}{|c}{\cellcolor[HTML]{FFFFFF}{\color[HTML]{000000} Teacher evaluation}}                                                                                 \\ 
\multicolumn{2}{c|}{\multirow{-2}{*}{\cellcolor[HTML]{FFFFFF}{\color[HTML]{000000} Models}}}                                  & {\cellcolor[HTML]{FFCCC9} MAE(graph)}       &  {\cellcolor[HTML]{CBCEFB} MAE(EC)}          & {\cellcolor[HTML]{FFFC9E} MAE(PC)}          & {\cellcolor[HTML]{FFCCC9} MAE(graph)}       &  {\cellcolor[HTML]{CBCEFB} MAE(EC)}          & {\cellcolor[HTML]{FFFC9E} MAE(PC)}          

\\\hline\hline                                 
\cellcolor[HTML]{FFFFFF}                                                  & {\color[HTML]{14A4C8} $\mathcal{L}_{a}$} & 0.11913                                 & {\color[HTML]{14A4C8} \textbf{0.00972}} & {\color[HTML]{14A4C8} \textbf{0.0008}}  & 0.13245                                 & {\color[HTML]{14A4C8} \textbf{0.00959}} & {\color[HTML]{14A4C8} \textbf{0.00078}} \\
\cellcolor[HTML]{FFFFFF}                                                  & {\color[HTML]{009901} $\mathcal{L}_{b}$} & 0.10437                                 & 0.01332                                 & {\color[HTML]{009901} \textbf{0.00108}} & {\color[HTML]{009901} \textbf{0.10438}} & 0.01356                                 & 0.00113                                 \\
\multirow{-3}{*}{\cellcolor[HTML]{FFFFFF}Ablated (augmented + real data)} & {\color[HTML]{A411AA} $\mathcal{L}^{\star}$} & 0.48436                                 & 0.04074                                 & 0.00546                                 & 0.35087                                 & 0.0284                                  & 0.00377                                 \\\hline
\cellcolor[HTML]{FFFFFF}                                                  & {\color[HTML]{14A4C8} $\mathcal{L}_{a}$} & 0.10347                                 & 0.04136                                 & 0.00251                                 & 0.0987                                  & 0.01375                                 & 0.00119                                 \\
\cellcolor[HTML]{FFFFFF}                                                  & {\color[HTML]{009901} $\mathcal{L}_{b}$} & 0.10442                                 & 0.01454                                 & 0.00117                                 & 0.10442                                 & 0.01319                                 & 0.00106                                 \\
\multirow{-3}{*}{\cellcolor[HTML]{FFFFFF}Ablated (few-shot learning)}     & {\color[HTML]{A411AA} $\mathcal{L}^{\star}$} & 0.10442                                 & 0.01454                                 & 0.00118                                 & 0.09867                                 & 0.014                                   & 0.00121                                 \\\hline\hline
\cellcolor[HTML]{FFFFFF}                                                  & {\color[HTML]{14A4C8} $\mathcal{L}_{a}$} & 0.11572                                 & 0.00989                                 & 0.00081                                 & 0.12914                                 & 0.00969                                 & 0.00079                                 \\
\cellcolor[HTML]{FFFFFF}                                                  & {\color[HTML]{009901} $\mathcal{L}_{b}$} & {\color[HTML]{009901} \textbf{0.10434}} & {\color[HTML]{009901} \textbf{0.01318}} & 0.00109                                 & 0.10441                                 & 0.01374                                 & 0.00112                                 \\
\multirow{-3}{*}{\cellcolor[HTML]{FFFFFF}GmTE-Net}                        & {\color[HTML]{A411AA} $\mathcal{L}^{\star}$} & {\color[HTML]{A411AA} \textbf{0.10436}} & {\color[HTML]{A411AA} \textbf{0.01234}} & {\color[HTML]{A411AA} \textbf{0.00103}} & 0.11921                                 & {\color[HTML]{A411AA} \textbf{0.00956}} & {\color[HTML]{A411AA} \textbf{0.00078}} \\\hline
\cellcolor[HTML]{FFFFFF}                                                  & {\color[HTML]{14A4C8} $\mathcal{L}_{a}$} & {\color[HTML]{14A4C8} \textbf{0.10065}} & 0.01432                                 & 0.00124                                 & {\color[HTML]{14A4C8} \textbf{0.09826}} & 0.01446                                 & 0.00125                                 \\
\cellcolor[HTML]{FFFFFF}                                                  & {\color[HTML]{009901} $\mathcal{L}_{b}$} & 0.10439                                 & 0.01378                                 & 0.00111                                 & 0.10443                                 & 0.01428                                 & 0.00116                                 \\
\multirow{-3}{*}{\cellcolor[HTML]{FFFFFF}GmTE-Net$^\star$}                       & {\color[HTML]{A411AA} $\mathcal{L}^{\star}$} & 0.10439                                 & 0.01368                                 & 0.00111                                 & {\color[HTML]{A411AA} \textbf{0.09814}} & 0.01426                                 & 0.00125                        \\\hline\hline
\end{tabular}}
\caption{\emph{Functional connectomic data-based comparison of our GmTE-Net with its two ablated versions in terms of architecture (first two rows) and its two variants (last two rows). We highlight in bold, blue, green and purple the best performances got for each method when using $\mathcal{L}_{a}$, $\mathcal{L}_{b}$ and $\mathcal{L}^{\star}$ and loss functions, respectively.} }
\label{tab:2}
\end{table}

\begin{enumerate}

\item \emph{Ablated (augmented + real data).} Here, we use a single decoder with 4 stacked layers to predict the multi-trajectory brain evolution where we used both real and simulated data to train both teacher and student networks.

\item \emph{Ablated (few-shot learning).} Here, we use the same architecture, but we train the teacher on a few-shots from our dataset, and we train the student on the simulated brain graphs.

\item \emph{GmTE-Net and  GmTE-Net$^\star$}. In these versions, we train our proposed GmTE-Net with and without few-shot learning strategy, respectively.

\end{enumerate}

\begin{table}[]
\vspace{-15pt}
\centering
\scalebox{0.7}{
\begin{tabular}{|
>{\columncolor[HTML]{FFFFFF}}c |
>{\columncolor[HTML]{FFFFFF}}c |
>{\columncolor[HTML]{FFFFFF}}c |
>{\columncolor[HTML]{FFFFFF}}c |
>{\columncolor[HTML]{FFFFFF}}c |
>{\columncolor[HTML]{FFFFFF}}c |
>{\columncolor[HTML]{FFFFFF}}c |
>{\columncolor[HTML]{FFFFFF}}c |
>{\columncolor[HTML]{FFFFFF}}c |
>{\columncolor[HTML]{FFFFFF}}c |}
\hline
\cellcolor[HTML]{FFFFFF}                                                   & \cellcolor[HTML]{FFFFFF}                                            & \multicolumn{4}{c|}{\cellcolor[HTML]{FFFFFF}Student evaluation}                                                                             & \multicolumn{4}{c|}{\cellcolor[HTML]{FFFFFF}Teacher evaluation}                                                                             \\ \cline{3-10}
\rowcolor[HTML]{FFFFFF} 
\multirow{-2}{*}{\cellcolor[HTML]{FFFFFF}Models}                           & \multirow{-2}{*}{\cellcolor[HTML]{FFFFFF}Losses} & Timepoints & {\cellcolor[HTML]{FFCCC9} MAE(graph)}                  & {\cellcolor[HTML]{CBCEFB} MAE(EC)}          & {\cellcolor[HTML]{FFFC9E} MAE(PC)}                            & Timepoints &{\cellcolor[HTML]{FFCCC9} MAE(graph)}                  & {\cellcolor[HTML]{CBCEFB} MAE(EC)}          & {\cellcolor[HTML]{FFFC9E} MAE(PC)}                              \\ \hline                               
\cellcolor[HTML]{FFFFFF}                                                   & \cellcolor[HTML]{FFFFFF}{\color[HTML]{14A4C8} }                     & $t_0$         & 0.335965                                 & {\color[HTML]{14A4C8} 0.036643}          & 0.00511                                  & $t_0$         & 0.291466                                 & 0.03059                                  & 0.004119                                 \\
\hline
\end{tabular}}
\caption{\emph{Morphological data-based comparison of our GmTE-Net with its two ablated versions in terms of architecture (first two rows) and its two variants (last two rows).} The best performances for each timepoint are highlighted in blue, green and purple when using $\mathcal{L}_{a}$, $\mathcal{L}_{b}$ and $\mathcal{L}^{\star}$ and loss functions, respectively. We further highlight in bold the best performance across four timepoints (i.e., average).} 
\label{tab:11}
\end{table}

% \usepackage{multirow}

\begin{table}[]
\centering
\hspace{-15pt}
\scalebox{0.7}{
\begin{tabular}{|
>{\columncolor[HTML]{FFFFFF}}c |
>{\columncolor[HTML]{FFFFFF}}c |
>{\columncolor[HTML]{FFFFFF}}c |
>{\columncolor[HTML]{FFFFFF}}c |
>{\columncolor[HTML]{FFFFFF}}c |
>{\columncolor[HTML]{FFFFFF}}c |
>{\columncolor[HTML]{FFFFFF}}c |
>{\columncolor[HTML]{FFFFFF}}c |
>{\columncolor[HTML]{FFFFFF}}c |
>{\columncolor[HTML]{FFFFFF}}c |}
\hline
\cellcolor[HTML]{FFFFFF}                                                   & \cellcolor[HTML]{FFFFFF}                                            & \multicolumn{4}{c|}{\cellcolor[HTML]{FFFFFF}{\color[HTML]{000000} Student evaluation}}                                                      & \multicolumn{4}{c|}{\cellcolor[HTML]{FFFFFF}{\color[HTML]{000000} Teacher evaluation}}                                                      \\ \cline{3-10} 
\multirow{-2}{*}{\cellcolor[HTML]{FFFFFF}Models}                           & \multirow{-2}{*}{\cellcolor[HTML]{FFFFFF}Losses}                    & Timepoints & {\cellcolor[HTML]{FFCCC9} MAE(graph)}                  & {\cellcolor[HTML]{CBCEFB} MAE(EC)}          & {\cellcolor[HTML]{FFFC9E} MAE(PC)}                            & Timepoints &{\cellcolor[HTML]{FFCCC9} MAE(graph)}                  & {\cellcolor[HTML]{CBCEFB} MAE(EC)}          & {\cellcolor[HTML]{FFFC9E} MAE(PC)}                               \\ \hline
\cellcolor[HTML]{FFFFFF}                                                   & \cellcolor[HTML]{FFFFFF}{\color[HTML]{14A4C8} }                     & $t_0$         & 0.121212                                 & 0.011029                                 & 0.000903                                 & $t_0$         & 0.125928                                 & 0.010554                                 & 0.000862                                 \\
\hline
\end{tabular}}
\caption{\emph{Functional data-based comparison of our GmTE-Net with its two ablated versions in terms of architecture (first two rows) and its two variants (last two rows).} The best performances for each timepoint are highlighted in blue, green and purple when using $\mathcal{L}_{a}$, $\mathcal{L}_{b}$ and $\mathcal{L}^{\star}$ and loss functions, respectively. We further highlight in bold the best performance across four timepoints (i.e., average).} 
\label{tab:22}
\end{table}

To evaluate our proposed GmTE-Net, we compute for each timepoint the MAE and the centrality scores (i.e., eigenvector (EC) and PageRank (PC) centralities metrics) between the ground truth and the predicted graphs. Mainly, we evaluate both teacher and student networks. Then, we report the results for each timepoint in \textbf{Table}~\ref{tab:11} and \textbf{Table}~\ref{tab:22} and we report the average prediction results across four timepoints in \textbf{Table}~\ref{tab:1} and \textbf{Table}~\ref{tab:2}. Specifically, \textbf{Table}~\ref{tab:1} shows the significant outperformance of GmTE-Net$^\star$ over the baseline methods when tested on morphological brain graphs. This demonstrates the breadth of our few-shot learning strategy in accurately training the teacher network. Thus, we can confirm the high-quality knowledge transferred from the teacher to the student network. Our GmTE-Net$^\star$ trained with the proposed global and local loss functions significantly outperformed its variants when using other losses ($p < 0.05$ using two-tailed paired t-test), which highlights the impact of our proposed topology-aware distillation loss in accurately learning the topology of the brain graphs predicted from the teacher network. While GmTE-Net$^\star$ outperformed other comparison methods when evaluated on the morphological dataset, it ranked second-best in MAE(graph), MAE(EC), and MAE (PC), following GmTE-Net when evaluated on functional brain graphs as displayed in \textbf{Table}~\ref{tab:2}. This shows that shifting the size of GCN layers to predict the brain graphs with a different resolution from the base one is not sufficient for getting an accurate prediction. Thereby, as a future research direction, we will include a domain alignment component in our architecture which will boost the prediction of brain graphs with different resolutions \cite{pilanci2020domain,redko2020survey}.

% %% ***************************************************************************** %%
\section{Conclusion}
% %% ***************************************************************************** %%
We proposed GmTE-Net, the first teacher-student framework tailored for predicting baby brain graph multi-trajectory evolution over time from a single neonatal observation. Our key contributions consist in: (i) designing a graph autoencoder able to chart out multiple trajectories with shifting graph resolutions from a single learned shared embedding of a testing neonatal brain graph, (ii) circumventing the learning on a few brain graphs such that we train the teacher network solely on a few shots and the student network on simulated brain graphs, and (iii) introducing a local topology-aware distillation loss to force the student model to preserve the node structure of the predicted graphs from teacher network. For the first time, we take the baby connectomics field one step further into foreseeing realistic baby brain graphs that may help to better elucidate how infant brain disorders develop. As a future direction, we will incorporate a domain alignment module into each decoder in our framework so that we reduce the domain fracture resulting in the distribution difference across brain modalities \cite{redko2020survey}.

\section{Acknowledgements}

This work was funded by generous grants from the European H2020 Marie Sklodowska-Curie action (grant no. 101003403, \url{http://basira-lab.com/normnets/}) to I.R. and the Scientific and Technological Research Council of Turkey to I.R. under the TUBITAK 2232 Fellowship for Outstanding Researchers (no. 118C288, \url{http://basira-lab.com/reprime/}). However, all scientific contributions made in this project are owned and approved solely by the authors. A.B is supported by the same TUBITAK 2232 Fellowship.

% %% ***************************************************************************** %%
\section{Supplementary material}
% %% ***************************************************************************** %%

We provide three supplementary items for reproducible and open science:

\begin{enumerate}
	\item A 5-mn YouTube video explaining how our prediction framework works on BASIRA YouTube channel at \url{https://youtu.be/rXH9K6a7OAI}.
	\item GmTE-Net code in Python on GitHub at \url{https://github.com/basiralab/GmTE-Net}. 
	%\item A GitHub video code demo on BASIRA YouTube channel at \url{https://youtu.be/2zKle7GzrIM}. 
\end{enumerate}

%\newpage
%%%%%%%%%%%%%%%%%%%%%%%%%%%%%%%%%%%%%%%%%%%%%%%%%%%%%%%%%%%%%%%%%%%%%%%%%%%%%%%%%%%%%%%%%%%%%%%%%%%%%%%%%%%%
\bibliography{Biblio3}
\bibliographystyle{splncs}
\end{document}